\documentclass[10pt,conference]{IEEEtran}
\IEEEoverridecommandlockouts
\usepackage{cite}
\usepackage{graphicx}
\usepackage{amsmath,amssymb}
\usepackage{booktabs}
\usepackage{multirow}
\usepackage{tabularx}
\usepackage{listings}
\usepackage{xcolor}
\usepackage{url}
\usepackage{enumitem}
\usepackage{algorithm}
\usepackage{algpseudocode}

\algrenewcommand\algorithmiccomment[1]{%
  \hfill$\triangleright$~\parbox[t]{0.44\linewidth}{\raggedright\footnotesize #1}%
}

\usepackage{tikz}
\usetikzlibrary{shapes,arrows,positioning,fit,calc,shadows.blur,backgrounds}
\usepackage{fontawesome5}
\usepackage[colorlinks=true, linkcolor=blue, citecolor=blue, urlcolor=blue]{hyperref}

\definecolor{codegreen}{rgb}{0,0.6,0}
\definecolor{codegray}{rgb}{0.5,0.5,0.5}
\definecolor{codepurple}{rgb}{0.58,0,0.82}
\definecolor{backcolour}{rgb}{0.95,0.95,0.92}

\lstdefinestyle{mystyle}{
    backgroundcolor=\color{backcolour},   
    commentstyle=\color{codegreen},
    keywordstyle=\color{blue},
    numberstyle=\tiny\color{codegray},
    stringstyle=\color{codepurple},
    basicstyle=\ttfamily\scriptsize,
    breakatwhitespace=false,         
    breaklines=true,                 
    captionpos=b,                    
    keepspaces=true,                 
    showspaces=false,                
    showstringspaces=false,
    showtabs=false,                  
    tabsize=2,
    frame=single,
    rulecolor=\color{black}
}

\title{Assertain: Automated Security Assertion Generation Using Large Language Models}

\author{\IEEEauthorblockN{Shams Tarek, Dipayan Saha, Khan Thamid Hasan, Sujan Kumar Saha,  Mark Tehranipoor, Farimah Farahmandi}
\IEEEauthorblockA{\textit{Department of Electrical and Computer Engineering, University of Florida, Gainesville, FL, USA}\\
\{shams.tarek, dsaha, khanthamidhasan, sujansaha\}@ufl.edu, \{tehranipoor, farimah\}@ece.ufl.edu}
\thanks{We thank the U.S. National Science Foundation (NSF) for support through CAREER Award No. 2339971.}
}
\newcommand{\papernote}{This paper will be presented at the 35th Microelectronics Design and Test Symposium (IEEE MDTS 2026)}

\makeatletter
\def\ps@IEEEtitlepagestyle{%
  \def\@oddhead{\parbox{\textwidth}{\centering \textbf{\papernote}}\hfill}%
  \def\@oddfoot{}%
  \def\@evenfoot{}%
}
\makeatother
\begin{document}

\maketitle

\begin{abstract}
The increasing complexity of modern system-on-chip designs amplifies hardware security risks and makes manual security property specification a major bottleneck in formal property verification. This paper presents \emph{\textit{Assertain}}, an automated framework that integrates RTL design analysis, Common Weakness Enumeration (CWE) mapping, and threat model intelligence to automatically generate security properties and executable SystemVerilog Assertions. \emph{\textit{Assertain}} leverages large language models with a self-reflection refinement mechanism to ensure both syntactic correctness and semantic consistency. Evaluated on 11 representative hardware designs, \emph{\textit{Assertain}} outperforms \textit{GPT-5} by 61.22\%, 59.49\%, and 67.92\% in correct assertion generation, unique CWE coverage, and architectural flaw detection, respectively. These results demonstrate that \emph{\textit{Assertain}} significantly expands vulnerability coverage, improves assertion quality, and reduces manual effort in hardware security verification.
\end{abstract}

\section{Introduction}
The increasing complexity and interconnectivity of modern system-on-chip (SoC) designs have significantly amplified hardware security challenges. Undetected vulnerabilities at the hardware level can result in severe consequences, including data breaches, financial loss, and system malfunction. Traditional verification flows, which largely depend on manual code inspection and simulation-based testing, struggle to scale with growing design complexity and to adapt to rapidly evolving threat landscapes. In this context, formal property verification (FPV) has emerged as one of the most rigorous techniques for validating design behavior against security requirements.

FPV enables exhaustive reasoning over all possible execution paths and provides mathematical guarantees of correctness when security properties are satisfied \cite{farahmandi2018formal}. Despite its theoretical strength, assertion-based verification (ABV), the primary mechanism used in FPV, remains difficult to scale in practice due to its heavy reliance on manual property specification. Writing security-centric properties and SystemVerilog assertions (SVAs) requires deep domain expertise, intimate knowledge of the design, and awareness of evolving threat models. This manual process is time-consuming, error-prone, and often biased toward verifying known or anticipated requirements, leaving subtle architectural weaknesses and emergent vulnerabilities undiscovered. As SoCs continue to grow in size and complexity and as threat models become more sophisticated, traditional ABV workflows increasingly struggle to keep pace, widening the gap between verification capability and security assurance needs.

Recent advances in large language models (LLMs) have demonstrated strong capabilities in understanding natural language, synthesizing code, and reasoning across multiple abstraction levels, making them promising candidates for assisting hardware security verification tasks. Early efforts have successfully applied LLMs to security asset identification \cite{hasan2026lasset}, threat modeling and test plan generation \cite{saha2025threatlens}, vulnerability insertion and detection \cite{saha2025sv, saha2024empowering, tarek2025socurellm, tarek2025bugwhisperer}, and Trojan detection \cite{chaudhuri2024spiced}. These successes highlight the potential of LLMs to reduce human effort, improve scalability, and expand coverage in security verification workflows.

However, existing LLM-based approaches for security property and assertion generation remain limited in several key aspects. Many techniques lack explicit threat awareness, do not systematically map design behaviors to standardized vulnerability classes, and suffer from hallucinated or context-inconsistent assertions that are not executable by formal verification tools. To address these limitations, this work introduces \emph{\textit{Assertain} }, an automated security property and SystemVerilog Assertion generation framework that unifies structural RTL analysis with threat model intelligence through design-aware common weakness enumeration (CWE) mapping.

\section{Related Work}
\label{background}

Recent advances in generative AI have led to a growing body of research that leverages LLMs to generate SystemVerilog Assertions (SVAs). Early frameworks such as \textit{AssertLLM} \cite{assertllm}, \textit{Chiraag} \cite{chirag}, and \textit{LAAG-RV} \cite{laag} demonstrated the feasibility of synthesizing assertions from natural-language functional specifications using few-shot prompting. Building on this direction, \textit{AssertionForge} \cite{bai2025assertionforge} unified design specifications and RTL into a knowledge graph to improve functional accuracy. While \textit{OpenAssert} \cite{menon2025openassert} addressed intellectual property (IP) security through fine-tuned open-source models for safe local deployment, \textit{FVDebug} \cite{bai2025fvdebug} focused on automating formal verification failure debugging by converting counterexample traces, RTL, and specifications into structured causal graphs. Despite these advances, quantitative evaluations from \textit{AssertionBench} \cite{pulavarthi2025assertionbench} highlight a critical limitation: even with sophisticated prompting strategies, standard LLMs exhibit a performance ceiling of approximately 40\% assertion validity, largely due to hallucinated signals and insufficient design context. Although existing approaches effectively reduce the semantic gap between specifications and RTL or address privacy and deployment concerns, they are primarily tailored to functional verification tasks.

Security property and assertion generation have received comparatively limited attention. Kande et al. \cite{kande2024security} evaluated the ability of LLMs to generate security-centric SVAs directly from natural language prompts. LASP \cite{ayalasomayajula2024lasp} later proposed an LLM-driven framework that extracts security assets from RTL and specifications to generate and catalog tailored security properties. More recently, Reddy et al. \cite{ankireddy2025lasa} introduced LASA, an LLM and retrieval-augmented generation-based framework that produces non-vacuous security properties and SVAs from SoC specifications through iterative refinement using formal verification feedback. In contrast to these approaches, which lack explicit threat modeling, design-aware CWE mapping, and systematic hallucination mitigation, our work integrates threat-aware reasoning with design contextualized vulnerability modeling to generate correct, relevant, and robust security-centric SVAs

\section{Methodology}

The core objective of \emph{\textit{Assertain} } is to automate the translation of high-level threat intent and low-level design artifacts into rigorous formal verification artifacts. 

\subsection{Formal Problem Formulation}
Let $D$ represent the RTL design space and $T$ denote the set of user-defined threat vectors. We use the CWE ontology, denoted as $K_{\text{CWE}}$, as the semantic bridge between threats and design structures. Our goal is to derive a mapping function $\mathcal{F}: (D, T) \rightarrow \mathcal{A}_{\text{final}}$, where $\mathcal{A}_{\text{final}}$ is a set of SVAs such that:
\begin{itemize}
    \item Syntactic Validity: $\forall a \in \mathcal{A}_{\text{final}}$, $a$ is compilable in SystemVerilog.
    \item Semantic Consistency: The signals referenced in $a$ exist in $D$ ($Signals(a) \subseteq Signals(D)$).
    \item Security Relevance: Each $a$ addresses a security weakness $w \in K_{\text{CWE}}$ derived from both $D$ and $T$.
\end{itemize}

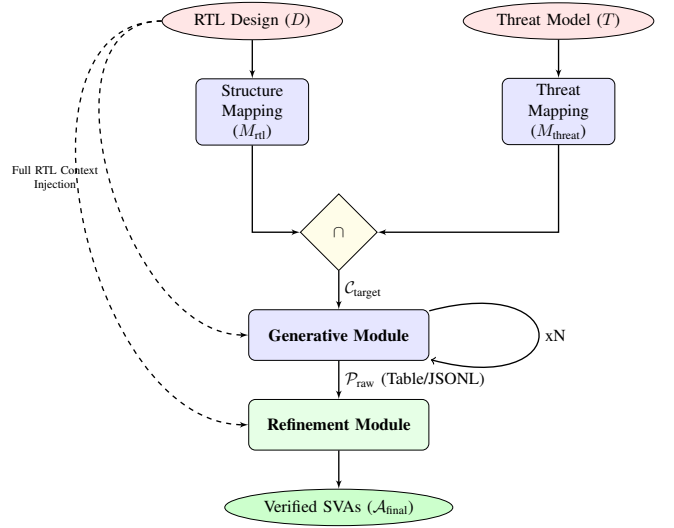
\begin{figure}[t]
\centering
\resizebox{1.0\columnwidth}{!}{
\begin{tikzpicture}[node distance=1.2cm, auto,
    block/.style={rectangle, draw, fill=blue!10, text width=6em, text centered, rounded corners, minimum height=3em},
    cloud/.style={draw, ellipse, fill=red!10, node distance=2.5cm, minimum height=2em},
    decision/.style={diamond, draw, fill=yellow!10, text width=4em, text centered, inner sep=0pt},
    line/.style={draw, -latex', thick},
    container/.style={draw, rectangle, dashed, inner sep=0.5em, fill=gray!5, rounded corners}]

    \node [cloud] (rtl) {RTL Design ($D$)};
    \node [cloud, right=of rtl] (threat) {Threat Model ($T$)};
    
    \node [block, below=0.8cm of rtl] (map1) {Structure Mapping ($M_{\text{rtl}}$)};
    \node [block, below=0.8cm of threat] (map2) {Threat Mapping ($M_{\text{threat}}$)};
    
    \node [decision, below=1cm of map1, xshift=1.8cm] (intersect) {$\cap$};
    
    \node [block, below=0.8cm of intersect, text width=10em] (gen) {\textbf{Generative Module}};
    \path [line] (gen) edge [loop right] node {xN} (gen);
    
    \node [block, below=0.8cm of gen, text width=10em, fill=green!10] (refine) {\textbf{Refinement Module}};
    
    \node [cloud, below=0.8cm of refine, fill=green!20] (sva) {Verified SVAs ($\mathcal{A}_{\text{final}}$)};

    \path [line] (rtl) -- (map1);
    \path [line] (threat) -- (map2);
    \path [line] (map1) |- (intersect);
    \path [line] (map2) |- (intersect);
    \path [line] (intersect) -- node[midway, right] {$\mathcal{C}_{\text{target}}$} (gen);
    \path [line] (gen) -- node[midway, right] {$\mathcal{P}_{\text{raw}}$ (Table/JSONL)} (refine);
    \path [line] (refine) -- (sva);
    
    \draw [line, dashed] (rtl.west) to[out=180,in=180] node[left, text width=2cm, align=center, font=\scriptsize] {Full RTL Context Injection} (gen.west);
    \draw [line, dashed] (rtl.west) to[out=180,in=180] (refine.west);

\end{tikzpicture}
}
\caption{The workflow of the proposed \textit{Assertain} framework. The generative module is queried iteratively ($\times N$), emits a structured table that is serialized as JSONL, and the design context injection provides the full RTL source.}
\label{fig:framework_flow}
\end{figure}

The proposed \emph{\textit{Assertain} } framework achieves this through a three-phase architecture: Knowledge Mapping, Context-Aware Generation, and Refinement. The overview of the proposed framework is shown in Figure \ref{fig:framework_flow}.

\subsection{Phase I: Knowledge Mapping}
To ground the generation process in reality, we first align both the design execution context and the adversarial context to a standardized ontology.

\subsubsection{Structural Alignment ($M_{\text{rtl}}$)}
The framework first employs an LLM to classify the input RTL design into one of several predefined hardware categories (e.g., ``Memory Components'', ``Communication Interfaces'', ``Processor Subsystems''). The classifier is constrained to select exactly one category, and responses are normalized using closest-match resolution to handle minor wording deviations. This classification step allows \textit{\textit{Assertain} }  to map the design to a relevant set of CWEs using a static knowledge base, without requiring the LLM to hallucinate vulnerability mappings directly. Table \ref{tab:rtl2cwe} details these structural mappings ($M_{\text{rtl}}: D \rightarrow \mathcal{C}_{\text{struct}}$), which link architectural classes to their most representative weakness identifiers.

\begin{table}[ht]
\centering
\caption{RTL Design to CWE Mapping Knowledge Base}
\label{tab:rtl2cwe}
\resizebox{\columnwidth}{!}{%

\begin{tabular}{p{0.41\linewidth}p{0.51\linewidth}}
\toprule
                       RTL Design Category &                             Representative CWE-IDs \\
\midrule
             Basic Digital Building Blocks &                       1245, 1254, 1255, 1261, 1298 \\
               Combinational Logic Designs &                             1248, 1254, 1255, 1298 \\
                  Sequential Logic Designs &                       1245, 1264, 1261, 1250, 1298 \\
                         Memory Components & 1189, 1220, 1222, 1223, 1224, 1251, 1257, 1260,\\
                  Communication Interfaces & 319, 441, 1191, 1190, 1258, 1295, 1262, 1263 \\
                Security and Crypto Blocks & 319, 325, 1240, 1241, 1279, 1270, 1290, 1294 \\
                SoC Integration Components & 1189, 1193, 1311, 1316, 1317, 1318, 1190, 1209 \\
             Processor and Core Subsystems & 1252, 1281, 1303, 1342, 1420, 1421, 1422, 1423 \\
                     Peripheral Interfaces &           1262, 1263, 1299, 1189, 1258, 1295, 1220 \\
                 Verification-Only Designs &                   1053, 1059, 440, 441, 1296, 1295 \\
         Custom Accelerators and IP Blocks &     1189, 1256, 1276, 1278, 1357, 1298, 1300, 1329 \\
          Clock and Power Management Units &     1247, 1232, 1304, 1209, 1351, 1338, 1384, 1261 \\
Error Handling and Fault Tolerance Modules &                        1261, 1334, 1264, 440, 1300 \\
\bottomrule
\end{tabular}

}
\end{table}

\subsubsection{Threat Vector Integration ($M_{\text{threat}}$)}
Simultaneously, we process the threat model inputs provided by the user (e.g., ``Side Channel Attack'', ``Privilege Escalation''). A mapping function ($M_{\text{rtl}}: T \rightarrow \mathcal{C}_{\text{threat}}$) normalizes these threat strings and performs a direct lookup in a predefined mapping database (Table \ref{tab:tm2cwe}) to retrieve associated CWE identifiers. This ensures that abstract security goals are translated into concrete, verifiable weakness classes.

\begin{table}[ht]
\centering
\caption{Threat model to CWE mapping used by \textit{Assertain} }
\label{tab:tm2cwe}
\resizebox{\columnwidth}{!}{%
\begin{tabular}{p{0.31\linewidth}p{0.61\linewidth}}
\toprule
               Threat Model &                             Representative CWE-IDs \\
\midrule
        Information Leakage & 203, 226, 319, 440, 441, 1243, 1246, 1258, 1295 \\
        Side Channel Attack & 1255, 1300, 1247, 1254, 1264, 1303, 1342, 1420 \\
     Fault Injection Attack & 1247, 1261, 1298, 1319, 1332, 1334, 1338, 1351 \\
Unauthorized Access control & 1190, 1191, 1244, 1242, 1247, 1256, 1257, 1262 \\
    Improper Access control & 1189, 1209, 1220, 1221, 1222, 1223, 1224, 1231 \\
          Denial of Service & 440, 1245, 1248, 1250, 1251, 1261, 1298, 1334 \\
     Confidentiality Attack & 226, 319, 1240, 1241, 1243, 1246, 1252, 1253 \\
       Privilege Escalation & 1189, 1193, 1190, 1191, 1209, 1221, 1242, 1252 \\
\bottomrule
\end{tabular}

}
\end{table}

\subsubsection{Context Intersection}
The target search space $\mathcal{C}_{\text{target}}$ is defined as the intersection of structurally implied weaknesses and threat-derived weaknesses:
\begin{equation}
\mathcal{C}_{\text{target}} = \mathcal{C}_{\text{struct}} \cap \mathcal{C}_{\text{threat}}
\end{equation}
This intersection ensures that we generate properties that are strictly relevant to the confirmed attack surface of the specific hardware implementation, reducing noise.
\begin{algorithm}[!ht]
\small
\caption{\textit{Assertain}  Generation \& Refinement Process}
\label{alg:threat2sva}
\begin{algorithmic}[1]
\raggedright
\Require Design Context $D$, Threat Set $T$, Knowledge Base $K_{\text{CWE}}$
\Ensure Set of Verified Assertions $\mathcal{A}_{\text{final}}$

\Statex \textbf{\textit{Stage 1: Semantic Alignment}}
\State $\mathcal{C}_{\text{struct}} \leftarrow \Call{MapStructure}{D, K_{\text{CWE}}}$
\State $\mathcal{C}_{\text{threat}} \leftarrow \Call{MapThreats}{T, K_{\text{CWE}}}$
\State $\mathcal{C}_{\text{target}} \leftarrow \mathcal{C}_{\text{struct}} \cap \mathcal{C}_{\text{threat}}$ \Comment{Target Intersection}

\Statex \textbf{\textit{Stage 2: Generative Synthesis}}
\State $\mathcal{P}_{\text{raw}} \leftarrow \emptyset$
\For{each weakness $w \in \mathcal{C}_{\text{target}}$}
    \State $\Phi \leftarrow \Call{BuildHybridPrompt}{D, w}$
    \For{$i \leftarrow 1$ to $N$} \Comment{Iterative Generation}
        \State $R \leftarrow \Call{LLMGen}{\Phi}$ \Comment{Returns a structured Markdown table}
        \State $\mathcal{T}_{\text{triplets}} \leftarrow \Call{ParseTable}{R}$ 
        \State \Comment{Schema check; keep numeric CWE rows; add CWE titles; extract $(Desc, Prop_{\text{NL}}, SVA_{\text{code}}, Tags)$}
        \State $\mathcal{P}_{\text{raw}} \leftarrow \mathcal{P}_{\text{raw}} \cup \mathcal{T}_{\text{triplets}}$
    \EndFor
\EndFor
\State $\mathcal{P}_{\text{raw}} \leftarrow \Call{Unique}{\mathcal{P}_{\text{raw}}}$ \Comment{Deduplication}
\State $\Call{SerializeJSONL}{\mathcal{P}_{\text{raw}}}$ \Comment{Staging for refinement}

\Statex \textbf{\textit{Stage 3: Self-Reflection Refinement}}
\State $\mathcal{A}_{\text{final}} \leftarrow \Call{LLMRefine}{\mathcal{P}_{\text{raw}}, D}$ \Comment{Filter \& emit \texttt{systemverilog} .sva}

\State \Return $\mathcal{A}_{\text{final}}$
\end{algorithmic}
\end{algorithm}
\subsection{Phase II: Context-Aware Generation}
In this phase, we deploy an LLM for context-aware generation. For each identified weakness $w \in \mathcal{C}_{\text{target}}$, we construct a prompt ($\Phi$) that encapsulates three layers of context:
\begin{enumerate}
    \item Role Context: Establishes the LLM as an expert hardware security engineer.
    \item Vulnerability Context: Provides the definition, common consequences, and demonstrative examples of weakness $w$.
    \item Design Context: Injects the full RTL source verbatim (single-file input in the current implementation), including module headers, signal declarations, and logic. The prompt explicitly asks the LLM to infer the top-level module and hierarchy to ground generation, without pre-filtering or slicing.
\end{enumerate}
The model is instructed to produce a structured triplet for each property: a \textit{Security Scenario} describing the attack vector, a \textit{Natural Language Property} formally stating the requirement, and a \textit{Candidate SVA} implementing the check.
The prompt explicitly encourages multiple scenarios and properties per CWE to maximize coverage.

To maximize the coverage of potential security scenarios and avoid the limitations of deterministic generation, we query the LLM iteratively ($N$ times, a configurable hyperparameter) for each target weakness. The results from all iterations are parsed and aggregated into a raw property set $\mathcal{P}_{\text{raw}}$, with identical properties (based on scenario and SVA code) automatically deduplicated. This ensemble approach mitigates the stochastic nature of LLMs and ensures a more comprehensive exploration of the state space.

To enable deterministic parsing, the model is constrained to return a strict markdown table with the columns \texttt{CWE ID | Security Scenario | NL Security Properties | SVAs}. Functional properties are encoded with \texttt{CWE ID = 0}. During parsing, only rows with numeric CWE identifiers are retained, and each retained row is augmented with its CWE title before being serialized as JSONL for the refinement stage.

\begin{table*}[t]
\scriptsize
\centering
\caption[Security Property Generation (Merged)]{Baseline vs. \textit{Assertain}: assertion correctness, CWE coverage, and architectural flaw detection.}
\label{tab:property_generation_merged}

\setlength{\tabcolsep}{3pt}
\renewcommand{\arraystretch}{1.15}

\resizebox{\textwidth}{!}{%
\begin{tabular}{@{}l*{10}{c}@{}}
\toprule
Design
& \multicolumn{2}{c}{\# Correct Assertions}
& \multicolumn{2}{c}{\% Syntactically Correct}
& \multicolumn{2}{c}{\% Functionally Correct}
& \multicolumn{2}{c}{\# Unique CWE}
& \multicolumn{2}{c}{\# Arch. Flaws Detected} \\
\cmidrule(lr){2-3}
\cmidrule(lr){4-5}
\cmidrule(lr){6-7}
\cmidrule(lr){8-9}
\cmidrule(lr){10-11}
& Baseline & \textit{Assertain} & Baseline & \textit{Assertain} & Baseline & \textit{Assertain} & Baseline & \textit{Assertain} & Baseline & \textit{Assertain} \\
\midrule
\texttt{Door\_Lock}              & 12 & 27 & 73  & 89  & 73  & 89  & 7  & 11 & 0 & 2  \\
\texttt{DMI\_JTAG}               & 12 & 30 & 92  & 97  & 92  & 97  & 6  & 11 & 0 & 2  \\
\texttt{Password\_Verification}  & 12 & 21 & 92  & 100 & 92  & 100 & 4  & 9  & 0 & 2  \\
\texttt{CSR\_Module}             & 23 & 46 & 100 & 100 & 100 & 100 & 11 & 15 & 8 & 10 \\
\texttt{AES\_Buggy}              & 30 & 31 & 100 & 100 & 100 & 100 & 11 & 6  & 6 & 7  \\
\texttt{SoC\_Subsystem}          & 33 & 32 & 100 & 88  & 100 & 88  & 7  & 15 & 7 & 8  \\
\texttt{MIPS}                    & 27 & 44 & 100 & 93  & 100 & 93  & 6  & 15 & 8 & 15 \\
\texttt{UART}                    & 21 & 33 & 100 & 100 & 100 & 100 & 9  & 7  & 4 & 7  \\
\texttt{Store\_Unit}             & 22 & 50 & 100 & 100 & 100 & 100 & 8  & 15 & 9 & 15 \\
\texttt{MMU}                     & 18 & 42 & 100 & 100 & 100 & 100 & 5  & 15 & 6 & 15 \\
\texttt{I2C}                     & 35 & 39 & 100 & 95  & 100 & 95  & 5  & 7  & 5 & 6  \\
\bottomrule
\end{tabular}%
}
\end{table*}

\subsubsection{Prompt Engineering \& Rule-Based Generation}
To ensure the generation of high-quality, compilable, and semantically valid SVAs, the prompt ($\Phi$) enforces a comprehensive set of syntactical, behavioral, and structural rules. These rules act as prompt-level integrity checks (signal binding, temporal correctness, non-vacuity/logical soundness, behavioral-spec perspective, and clock-event hygiene) to prevent common formal verification pitfalls.

\begin{itemize}[leftmargin=*, labelsep=0.6em, itemsep=2pt, topsep=2pt]
    \item Scope and Perspective: The LLM is directed to generate properties for both internal module logic and interface protocols between connected modules. Crucially, properties must be written from the \textit{behavioral specification perspective}, ensuring they verify intended functionality rather than merely mirroring the implementation.
    
    \item Syntax and Structure: 
    \begin{itemize}[leftmargin=1.2em, labelsep=0.4em, itemsep=1pt, topsep=1pt]
        \item Adherence to the \texttt{antecedent |-> consequent} format is mandatory.
        \item The antecedent must be a meaningful enable condition (e.g., \texttt{valid}, \texttt{req}) rather than a trivial state.
        \item The consequent must reflect the expected behavior under the antecedent.
        \item Named properties are enforced to enhance reusability and readability (e.g., \texttt{property p\_secure\_handshake}).
        \item Implication operators (\texttt{|->}, \texttt{|=>}) are used for conditional and temporal relationships.
    \end{itemize}

    \item Strict Timing and Reset Logic:
    \begin{itemize}[leftmargin=1.2em, labelsep=0.4em, itemsep=1pt, topsep=1pt]
        \item Synchronous Compliance: The prompt enforces a critical distinction in implication operators. The non-overlapping operator (\texttt{|=>}) must be used for effects occurring in the \textit{next clock cycle}, while the overlapping operator (\texttt{|->}) is reserved for combinatorial or same-cycle effects; violations are treated as critical errors.
        \item Reset Handling: All assertions must include an appropriate \texttt{disable iff} reset condition to prevent spurious firings during reset phases, with \texttt{rst\_n} as the default convention unless a design-specific reset is provided.
        \item Cycle-Accuracy: Preference is given to cycle-accurate temporal assertions over purely combinatorial checks, avoiding reliance on physical artifacts like glitches or inertial delays.
        \item Clocking Hygiene: The signal used in the clocking event must not be reused inside the property body, preventing self-referential checks.
    \end{itemize}

    \item Signal Integrity and Binding: 
    \begin{itemize}[leftmargin=1.2em, labelsep=0.4em, itemsep=1pt, topsep=1pt]
        \item The LLM is strictly confined to use only existing RTL signals, preventing hallucinated names.
        \item Assertions must be bound to relevant signals; for instance, a confidentiality property must inspect actual data lines (e.g., \texttt{wdata}, \texttt{rdata}) rather than just control flags. This acts as an anti-hallucination guardrail.
    \end{itemize}

    \item Quality and Robustness:
    \begin{itemize}[leftmargin=1.2em, labelsep=0.4em, itemsep=1pt, topsep=1pt]
        \item Vacuous Property Prevention: The system explicitly forbids assertions that are trivially true. Checks must facilitate meaningful behavioral verification.
        \item Overlapping Pitfalls: To prevent ambiguity, edge-triggered sequences (using \texttt{\$rose}) are preferred to define clear start points for assertions.
        \item Invariant Discipline: Unconditional properties over sequences are avoided unless explicitly verifying invariants.
        \item Logical Soundness: Properties must describe desirable correctness conditions without contradicting the design's primary function (i.e., verifying intended behavior, not impossible states).
        \item Built-in Efficiency: The use of SystemVerilog built-ins such as \texttt{\$rose}, \texttt{\$past}, \texttt{\$stable}, \texttt{\$isunknown}, and \texttt{\$onehot} is required for compact and accurate verification.
    \end{itemize}
\end{itemize}
\subsubsection{Functional Property Analysis}
Beyond its security-focused methodology, \textit{Assertain}  also applies a stringent Functional Property Analysis process. In this phase, the LLM must examine \textit{every module} line by line, reviewing key RTL elements, such as \texttt{always} blocks, finite state machines (FSMs), combinational logic assignments, handshakes, counters, and FIFOs, to infer the intended logical behavior.
The generation of these functional properties adheres to specific breadth and depth criteria:
\begin{itemize}[leftmargin=*, labelsep=0.6em, itemsep=2pt, topsep=2pt]
    \item Breadth: The analysis must cover all provided modules and their key interactions (interface protocols).
    \item Depth: Generated properties must capture complex behaviors such as temporal sequencing, coincidence, gating, error-handling, integrity, and liveness.
    \item Integration: Functional properties are translated into syntactically correct SVAs using the same strict rules as security properties. 
\end{itemize}
\subsection{Phase III: Self-Reflection Refinement}
The direct outputs of probabilistic models frequently exhibit hallucinations, referring to signals that do not exist or using invalid syntax. To mitigate this issue, we introduce a self-reflection refinement stage.
The \textit{Refinement Module} receives two inputs: the raw candidate assertions ($\mathcal{P}_{\text{raw}}$) and the full RTL source code ($D$). It is instructed to perform a semantic derivation step:
The \textit{Refinement Module} takes as input the raw candidate assertions $\mathcal{P}_{\text{raw}}$ and the complete RTL source code $D$, and performs the following semantic derivation steps:
\begin{enumerate}[leftmargin=*, labelsep=0.6em, itemsep=2pt, topsep=2pt]
    \item Verification: For each candidate assertion $s \in \mathcal{P}_{\text{raw}}$, extract valid identifiers from the RTL (inputs, outputs, wires, regs, parameters, typedefs, enum values, packed-struct fields, and module-visible submodule ports) and verify that $Signals(s) \subseteq Signals(D)$.
    \item Filtering: Remove assertions that reference hallucinated or undefined identifiers (including undefined macros), while preserving valid typecasts, struct/enum field accesses, and module-scoped instances or types that are declared in the RTL.
    \item Synthesis: Emit the remaining assertions as a clean, compilable \texttt{.sva} module with a \texttt{timescale} directive, wrapper module, signal declarations (including typedefs/enums as needed), file headers, section annotations, and required macros. Each assertion includes structured scenario/NL comments and a meaningful \texttt{\$error} message.
\end{enumerate}
The refinement output is generated as a single \texttt{systemverilog} fenced code block and written verbatim as a standalone \texttt{.sva} file. When explicitly requested, the refiner can also synthesize an intricate property suite (14--20 properties, at least six with multi-cycle or cross-module dependencies) with grouped sections and optional \texttt{cover property} blocks for legal sequences. This process ensures that the final output $\mathcal{A}_{\text{final}}$ is syntactically robust and grounded in the actual design. The complete generation flow is detailed in Algorithm \ref{alg:threat2sva}.

In the reference implementation, intermediate artifacts are staged in a temporary workspace as JSONL, and the pipeline fails fast if no CWE intersection is found or if API credentials are missing. For reproducibility, all LLM calls are seeded and distinct models are used for classification (\textit{GPT-4o}), generation (\textit{GPT-5}), and refinement (\textit{GPT-4o}). The full RTL text is printed for traceability, and an orchestrator invokes the generator and refiner via framework component modules.

\section{Experiments and Results}
\subsection{Experimental Results}
In this experiment, \textit{GPT-5} was considered as an underlying model of the \textit{Assertain} framework. Table \ref{tab:property_generation_merged} compares the baseline LLM (\textit{GPT-5}) with the proposed framework, \textit{Assertain}, across a diverse set of 11 RTL designs. The evaluation focuses on three primary dimensions: (i) quantity and correctness of generated assertions, (ii) CWE coverage, and (iii) detection of architectural flaws. From the table, it can be seen that \textit{Assertain} outperforms \textit{GPT-5} by 61.22\%, 59.49\%, and 67.92\% in terms of the number of correct assertion generations, unique CWE coverage, and architectural flaw detection, respectively.  
\subsubsection{Assertion Quantity and Correctness}
From Table \ref{tab:property_generation_merged}, it is shown that across nearly all designs, \textit{Assertain}  produces a higher number of correct security assertions, demonstrating its ability to capture security intent more comprehensively. For instance, in the \texttt{Store\_Unit} and \texttt{MMU} designs, \textit{Assertain}  more than doubles the correct assertion count compared to the baseline, highlighting improved coverage of memory protection and access control scenarios. Similarly, for control-intensive designs such as \texttt{Door\_Lock} and \texttt{DMI\_JTAG}, \textit{Assertain} generates substantially more correct SVAs that directly address access-control and privilege-escalation risks.

While the baseline achieved 100\% correctness on smaller or less complex designs, \textit{Assertain}  matched this performance in most cases. Some larger designs, such as \texttt{SoC\_Subsystem} and \texttt{MIPS}, showed slightly lower functional correctness (88\% and 93\%, respectively), reflecting the inherent difficulty of integration-heavy systems. Nonetheless, \textit{Assertain}  outperformed the baseline in nearly all scenarios in terms of both volume and reliability of assertions.

\subsubsection{CWE Coverage}
\textit{Assertain} consistently identified a broader range of vulnerabilities in most designs, with benchmarks such as \texttt{DMI\_JTAG}, \texttt{MMU}, and \texttt{MIPS} exhibiting more than twice as many unique CWE classes as the baseline. In terms of correct CWE mappings, \textit{Assertain}  also demonstrates broader coverage. Designs such as \texttt{CSR\_Module}, \texttt{SoC\_Subsystem}, and \texttt{MIPS} are associated with up to 15 relevant CWE identifiers, ensuring that generated properties are not only syntactically valid but also semantically aligned with security vulnerabilities and threat models. 

\subsubsection{Architectural Flaw Detection}
A major strength of \textit{Assertain}  lies in its ability to uncover architectural flaws. As shown in Table~\ref{tab:property_generation_merged}, the framework detected substantially more high-level issues than the baseline across nearly all benchmarks. For example, in the \texttt{MIPS} design, \textit{Assertain}  detected 15 flaws compared to 8 by the baseline. Similarly, in the \texttt{Store\_Unit} and \texttt{MMU}, the framework doubled the number of flaws detected (15 vs. 9 and 15 vs. 6, respectively). These findings highlight the framework's ability to capture nuanced vulnerabilities that are typically overlooked by monolithic LLM prompting.  


\subsection{Case Studies}
\label{sec:case_study}
\subsubsection{Selected Properties for \texttt{DMI\_JTAG}}
To demonstrate the practical efficacy of \textit{Assertain}, we deployed the framework on the \texttt{DMI\_JTAG} module of the PULP platform RISC-V debug infrastructure. This module implements the Debug Module Interface over JTAG and serves as a critical security boundary. Two representative properties generated by \textit{Assertain}  are shown in Listings~\ref{lst:sva_1295} and~\ref{lst:sva_441}.


\begin{lstlisting}[language=Verilog, caption={SVA for CWE-1295 on DMI\_JTAG: Enforcing Zero Data on Error. This property enforces a strict confidentiality invariant by ensuring that no residual data is exposed when the debug interface reports an error. By explicitly requiring data fields to be zeroed during busy or error states, it prevents stale information from being observable through error driven side channels. The assertion therefore captures a security critical architectural requirement that is typically absent from standard functional specifications.}, label={lst:sva_1295}]
property p_zero_data_on_error_capture_dbg;
  @(posedge tck_i) disable iff (!trst_ni) 
    capture_dr && dmi_access && (error_q == DMIBusy || error_dmi_busy) |=> dr_q[33:2] == '0;
endproperty
\end{lstlisting}

\begin{lstlisting}[language=Verilog, caption={SVA for CWE-441 on DMI\_JTAG: Validating Transaction Origin. This property enforces provenance integrity for core bound requests by requiring each transaction to be causally linked to a valid and error free JTAG update sequence. By constraining request issuance to legitimate protocol transitions, it prevents trusted logic from being abused as a confused deputy to execute unauthorized actions. The assertion thus captures a critical control path security invariant that is rarely made explicit in functional specifications.
}, label={lst:sva_441}]
property p_req_valid_preceded_by_update;
  @(posedge tck_i) disable iff (!trst_ni) 
    $rose(dmi_req_valid) |-> $past(dmi_access && update_dr && error_q == DMINoError);
endproperty
\end{lstlisting}


\subsubsection{Selected Properties for \texttt{UART}}
The \texttt{UART} subsystem integrates a UART, a DMA controller, and a debug bridge, introducing security risks related to unrestricted debug access and weak access control over configuration and memory-mapped operations. Using its standard pipeline, \textit{Assertain} identified the DMA and debug interfaces, mapped structural features to relevant CWE classes, and aligned them with the user-specified threat vector \textit{Improper Access Control}. 

\begin{lstlisting}[language=Verilog, caption={SVA for CWE-284: Confidentiality under Debug. When debug mode is active, sensitive configuration bits (DMA enable, priority) must not be exposed. The NL-Property states that sensitive fields must be masked or cleared during any active debug session.}, label={lst:uart_confidentiality}]
assert property (@(posedge clk) disable iff (!rst_n)
  (dbg_sel && dbg_en) |-> (csr_q.enable_dma == 1'b0 && csr_q.dma_prio == 3'h0));
\end{lstlisting}

\begin{lstlisting}[language=Verilog, caption={SVA for CWE-1244: No Sensitive Reads. Debug readback should never leak sensitive data. The NL-Property requires that debug output be masked to known benign constants under debug conditions.}, label={lst:uart_no_sensitive_read}]
assert property (@(posedge clk) disable iff (!rst_n)
  (dbg_sel && dbg_en) |-> (dbg_rdata == 32'hDEADBEEF || dbg_rdata == 32'hCAFEBABE));
\end{lstlisting}




\subsection{Architectural Gap Analysis}
Beyond standard functional verification, the \textit{Assertain} framework can perform architectural gap analysis by mining real-world security properties that reflect industry best practices. In traditional verification workflows, an assertion that fails during simulation is often dismissed as functionally incorrect or attributed to model hallucination because it contradicts the existing RTL implementation. In contrast, within the context of security verification, \textit{Assertain} intentionally generates assertions that act as negative tests, evaluating whether the design adheres to stronger security principles rather than merely conforming to its current behavior. As a result, assertion failures highlight architectural security gaps instead of functional bugs, exposing missing protections that could be exploited by adversaries. The following case studies illustrate this capability across different design modules.
\subsubsection{Case Study 1: Replay Attack Prevention}
During the evaluation of the \texttt{Password\_Verification} module, \textit{Assertain}  generated the following property to enforce a dynamic challenge-response protocol:
\begin{lstlisting}[language=Verilog, caption={Replay Attack Prevention}, label={lst:replay}]
// Scenario: Prevent predictable/reused challenges (back-to-back verifications)
// CWE-1241: Use of Predictable Algorithm in Random Number Generator
property p_no_reuse_back_to_back;
  @(posedge clk) disable iff (!rst_n)
  ($rose(isHashVerified) && $past($rose(isHashVerified)))
  |-> (storedHash != $past(storedHash));
endproperty
\end{lstlisting}
In the target RTL, \texttt{storedHash} remains static with no rotation mechanism, causing the assertion to fail. Rather than treating this as a functional mismatch, \textit{Assertain} identifies it as an architectural gap that exposes a replay attack risk due to the absence of challenge rotation or nonce generation.
\subsubsection{Case Study 2: Fault Tolerance and Self-Recovery}
In the \texttt{Door\_Lock} design, \textit{Assertain}  identified a reliability gap against Single Event shocks (SEUs) or fault injection. It generated the following property, asserting that if the FSM enters an undefined or illegal state (e.g., due to a bit-flip), it must automatically recover to a safe state in the next cycle:
\begin{lstlisting}[language=Verilog, caption={Fault Tolerance (SEU Recovery)}, label={lst:seu}]
// Scenario: SEU flips FSM into illegal encoding
// CWE-1245: Improper Finite State Machines (Fault Tolerance)
property p_seu_recover_illegal;
  @(posedge i_clk)
  (current_state == UNUSED_1 || current_state == UNUSED_2)
  |=> current_state == INITIAL_SET;
endproperty
\end{lstlisting}

The failure of this assertion highlights a gap in fault tolerance, where the system is fragile to physical attacks or environmental radiation that could trap the FSM in a deadlock or undefined state.

\subsubsection{Case Study 3: Role Isolation (Confused Deputy)}
For the \texttt{I2C} controller, the framework identified a privilege separation requirement and generated a security property enforcing mutual exclusion between the Master (controller) and Slave (target) roles:
\begin{lstlisting}[language=Verilog, caption={Role Isolation (Confused Deputy)}, label={lst:confused_deputy}]
// Scenario: Avoid appearing as both master and proxying slave at the same time
// CWE-441: Unintended Proxy or Intermediary ('Confused Deputy')
property p_no_slave_while_master;
  @(posedge clk) disable iff (!nReset)
  master_mode |-> !slave_act;
endproperty
\end{lstlisting}
Although the I\textsuperscript{2}C protocol supports multi-master configurations, a single device simultaneously acting as both master and slave can introduce state ambiguity and enable privilege escalation. If the RTL permits \texttt{master\_mode} and \texttt{slave\_act} to assert concurrently due to logic errors or race conditions, it constitutes a role isolation gap. \textit{Assertain} correctly identifies the need to explicitly interlock these roles to prevent Confused Deputy scenarios.

\subsubsection{Case Study 4: Glitch Resistance}
In the simple \texttt{Password\_\allowbreak Verification} module, \textit{Assertain} checked for robustness against voltage or clock glitches on the control line. It generated the following property, asserting that a verification signal lasting only a single clock cycle should be ignored as noise:

\begin{lstlisting}[language=Verilog, caption={Glitch Resistance}, label={lst:glitch}]
// Scenario: Resist single-cycle glitch on isHashVerified
// CWE-1247: Improper Protection Against Voltage and Clock Glitches
property p1247_no_single_cycle_accept;
  @(posedge clk) disable iff (!rst_n)
  (state==AUTH_CHECK_STATE && $rose(isHashVerified) && !$past(isHashVerified))
  |-> (!authenticationStatus);
endproperty
\end{lstlisting}
The original design was purely synchronous and trusted any high signal on the clock edge, effectively allowing a single-cycle glitch to trigger authentication. The failure of this property identified an architectural gap in glitch mitigation, flagging the need for a multi-cycle stability filter (debouncer) on critical security control inputs.

These examples demonstrate that the \textit{Assertain}  LLM does not merely verify if the implementation matches the specification, but evaluates if the specification itself is secure. By generating properties that fail on insecure designs, the model acts as an automated security architect, defining the ``to-be'' state of a hardened system.

\section{Conclusion}
\textit{Assertain} advances security property generation by correlating security weaknesses with threat models using CWE intelligence and applying a self-reflection refinement cycle. The framework generates executable SVAs that achieve higher functional accuracy and broader security coverage than a baseline LLM while reducing manual effort. Future work will expand the CWE knowledge base, incorporate retrieval augmented specification grounding, and evaluate the approach on full SoC platforms.


\bibliographystyle{IEEEtran}
\bibliography{GOMACTech_LaTeX}
\end{document}